\documentclass{pas}

\usepackage{natbib}
\usepackage{multirow}

\usepackage[nolist]{acronym}
\usepackage[range-units = single]{siunitx}

\begin{acronym}
\acro{LMXB}{low mass X-ray binary}
\acro{NS}{neutron star}
\acro{MCMC}{Markov-Chain Monte Carlo}
\acro{AMP}{accreting millisecond pulsar}
\acro{RMS}{root mean squared}
\end{acronym}

\DeclareUnicodeCharacter{2212}{-}

\newcommand{\ixpe}{{\it IXPE}}
\newcommand{\insight}{{\it Insight-HXMT}}
\newcommand{\srga}{SRGA~J144459.2$-$604207}
\newcommand{\sax}{SAX~J1808.4$-$3658}
\newcommand{\igr}{IGR~J17498$-$2921}
\newcommand{\gs}{GS~1826$-$24}
\newcommand{\zcno}{\ensuremath{Z_{\rm CNO}}}
\newcommand{\mns}{\ensuremath{M_{\rm NS}}}
\newcommand{\rns}{\ensuremath{R_{\rm NS}}}

\begin{document}

\lefttitle{Publications of the Astronomical Society of Australia}
\righttitle{Cambridge Author}

\jnlPage{1}{16}
\jnlDoiYr{2026}
\doival{10.1017/pasa.xxxx.xx}

\articletitt{Research Paper}

\title{Constraining neutron-star properties with ensembles of thermonuclear bursts:
application to SRGA~J144459.2$-$604207}

\author{\sn{Galloway}, \gn{Duncan K.}$^{1,2}$,
\sn{Jones}, \gn{Serena}$^3$, 
\sn{Waterson}, \gn{Luke}$^1$, 
\sn{Fu}, \gn{Tao}$^4$, 
\sn{Goodwin}, \gn{Adelle J.}$^3$,
and \sn{Li}, \gn{Zhaosheng}$^5$}

\affil{$^1$School of Physics \& Astronomy, Monash University, Clayton, VIC 3800, Australia and 
$^2$Institute for Globally-Distributed Research and Education (IGDORE); \\
$^3$
International Centre for Radio Astronomy Research, Curtin University, GPO Box U1987, Perth, WA 6845, Australia; 
$^4$ Key Laboratory of Stars and Interstellar Medium, Xiangtan University, Xiangtan 411105, Hunan, China; 
$^5$School of Science, Qingdao University of Technology, Qingdao 266525, China}

\corresp{D. K. Galloway, Email: duncan.galloway@monash.edu}

\citeauth{Galloway et al., Constraining neutron-star properties with ensembles of
thermonuclear bursts: application to SRGA J144459.2−604207. {\it Publications of the Astronomical Society of Australia} {\bf 00}, 1--16. https://doi.org/10.1017/pasa.xxxx.xx}

\history{(Received xx xx xxxx; revised xx xx xxxx; accepted xx xx xxxx)}

\begin{abstract}
Deducing the properties of the host neutron stars from thermonuclear (type-I) X-ray bursts remains a challenge, due to incomplete data, the large (multidimensional) parameter space, and dearth of suitable models and analysis tools. Here we describe further development of the {\sc beansp} package, used previously to analyse several burst samples. We present a new ``ensemble'' analysis mode, using the consistent and regular ``clocked'' bursting exhibited by some sources, that compares model predictions to epoch averages instead of individual events. This mode requires only one model evaluation per epoch, and so is much more efficient that the previous ``train'' mode. We apply the code to the best-known source exhibiting ``clocked'' bursting, GS~1826$-$24, using a grid of {\sc kepler} models pre-calculated for this purpose, and find good agreement with a previous study. 
We performed a number of experiments on simulated data, demonstrating that we can recover input parameters related to the burst ignition with reasonable accuracy, but less so for the system distance, emission anisotropy and neutron star mass and radius.
Finally, we assembled a set of 14 daily burst epochs covering the 2024 outburst of the accretion-powered millisecond pulsar \srga, and attempted to constrain the system properties of this object by comparing to {\sc settle} model predictions. 
We find reasonably good agreement between the observations and model predictions for a mildly sub-solar fuel composition, with H-fraction $X\simeq0.54$ and CNO metallicity $\zcno\simeq0.01$. However, we note that the inferred H-fraction is in excess of the limit of 0.4 established separately, and the adopted model may not provide sufficiently accurate predictions for this burst ignition regime. 
The inferred distance depends on assumptions about the system inclination and corresponding anisotropy of the persistent emission, and is likely in the range 6--11~kpc.
Future applications with more physically realistic models are a promising avenue for this and other sources with H-rich bursts.
\end{abstract}


\maketitle

\section{Introduction}
\label{sec:intro}

Type-I X-ray bursts are the observational signature of unstable thermonuclear burning on the surface of accreting \acp{NS}.
These events have been detected in approximately 120 Galactic sources\footnote{see \url{https://burst.sci.monash.edu/sources}}, and typically last from 10~s to a few minutes, while recurring on timescales of hours--days \cite[e.g.][]{minbar}.
The burst ignition conditions are primarily determined by the rate of mass accretion onto the neutron star, and the composition of the accreted fuel \cite[e.g.][]{gal21a}; the majority of systems have main-sequence mass donors and consequently accrete H-rich fuel. The composition of the fuel layer at ignition sets the properties of the bursts themselves, and the accreted H can be reduced or exhausted by steady burning prior to ignition. 

Despite decades of study, we remain largely ignorant of the accreted composition of most burst sources, which in turn contributes substantial uncertainty to the interpretation of their burst properties and behaviour. Additionally, the distance to most sources are rather poorly constrained, as is the impact of other systemic properties such as the inclination and the degree of emission anisotropy  arising from interactions of the outgoing radiation with the accretion disc \cite[e.g.][]{he16}.

The {\sc beansp} code has been used to determine \ac{NS} system parameters by comparing observations of thermonuclear bursts to numerical models, for several sources (\citealt{goodwin19c,gal24}; see also \citealt{gal06c}).
In the default burst ``train'' mode, {\sc beansp} uses a numerical ignition code, {\sc settle}, to generate a 
sequence of bursts in response to a mass accretion rate history inferred from persistent X-ray flux measurements of the target.
Each observed burst is matched to one of the simulated bursts, and an overall likelihood is calculated according to the agreement between the burst start times, fluences (integrated flux), and $\alpha$ (the ratio of the integrated persistent flux over the burst interval, to the burst fluence).
We then apply a \ac{MCMC} algorithm to explore the full parameter space and estimate posterior distributions for each of the burst and system parameters.

The {\sc beansp} code allows in principle full coverage of the parameter space, including marginalisation over several system parameters (e.g. distance, inclination) which might be considered ``nuisance'' in the context of measuring the neutron star properties (e.g. accreted composition, mass, radius). However, in it's original implementation there are several disadvantages. First, the code relies on detecting as many bursts as possible, closely spaced in time; this requirement favours uninterrupted observations of bursting sources, which are rarely achieved in practise. 
Second, as {\sc beansp} normally operates over intervals in which the accretion rate is not constant, the code uses an iterative method to find the burst recurrence time that matches the average accretion rate over that interval, according to the model and the current model parameters. This step can take several calls to the underlying ignition model to obtain a self-consistent combination of burst interval and averaged persistent flux.
To date, the largest burst sample that has been analysed with {\sc beansp} includes just 8 bursts (although likely spanning a sequence of 19, including events that fell in observation gaps; \citealt{gal24}).

This paper describes a modified approach taking advantage of `ensembles' of observed bursts with unambiguous recurrence times, to constrain the system parameters.
In Section \ref{sec:data} we describe the steps taken to assemble the data.
In Section \ref{sec:method} we present the alternate simulation method, and describe how the \ac{MCMC} runs were performed, also for the simulation tests.
In Section \ref{sec:results} we describe the results of application of the method to the burst sources \gs\ and \srga.
Finally in Section \ref{sec:disc} we interpret our results and provide a guide for future applications to other sources.

\section{Data and analysis}
\label{sec:data}

We adopted new and previously-analysed observations of thermonuclear bursts made by X-ray satellites including the {\it Rossi X-ray Timing Explorer}, {\it Insight-HXMT} and {\it IXPE}. As the new ``ensemble'' approach to matching burst observations requires regular, consistent bursts, we focused on a few systems that show such behaviour; the best known is \gs\ \cite[e.g.][]{gal03d}. These systems are often termed ``clocked bursters'' in 
{ published work \cite[e.g.][]{dohi25} }
although this term is a misnomer as such sources show regular (``clocked'') bursting behaviour only episodically \cite[see e.g.][]{chenevez16}.

We used an ensemble of bursts from \gs\ assembled previously by \cite{gal17a}, and compared against {\sc kepler} models by \cite{johnston20}. These data include measurements of the recurrence time $\Delta t$, integrated burst fluence $E_b$, and bolometric persistent flux $F_p$, over three epochs observed between 1998 and 2007. The sample data also contains estimates of $\alpha$ for each sequence of bursts, the ratio of integrated persistent (accretion) energy to burst fluence, a measure of the relative efficiency of the thermonuclear burning and accretion processes. This ``reference'' sample also includes burst lightcurves, but these were not used in the present work.

We also assembled a new set of data from the remarkable accretion-powered millisecond pulsar \srga, discovered by the Mikhail Pavlinsky ART-XC telescope on board the {\it Spectrum-Roentgen-
Gamma observatory} (SRG) when it went into outburst in 2024 \cite[]{molkov24}.
{\it NICER}\/ observations revealed 447.9~Hz pulsations, modulated on a 5.2~hr binary orbit \cite[]{ng24a}. That and other instruments that observed the source early in the outburst also revealed frequent, regular thermonuclear bursts without strong evidence for photospheric radius-expansion, unusual for accretion-powered millisecond pulsars \cite[e.g.][]{disalvo22}.
Based on the energetics and recurrence time of the bursts, \cite{fu25} derived an upper limit on the Hydrogen mass fraction of $X<0.4$. Detection of polarisation in the persistent emission led to constraints on the inclination \cite[]{papitto25}.

Our \srga\ data comes from 94 unique bursts observed by {\it Insight-HXMT} and {\it IXPE}, including 18 bursts observed by both instruments. For each element of our ``ensemble'' data set, we fit a linear model to the arrival time of each burst 
{ (including duplicate events\footnote{{ While this ``double counting'' of the duplicate bursts might unrealistically reduce the errors on the model fit parameters, in practise the uncertainties are dominated by systematic variations, likely arising from short-term variations in the accretion rate. }
})}
within daily epochs, to derive the daily mean recurrence time $\Delta t_i$ and uncertainty. 

Also within each epoch we calculated averages of the burst fluence $E_b$, persistent (bolometric) flux $F_p$ and $\alpha$ value.
For the \insight\ bursts, we used  tabulated values from \cite{fu25}. For \ixpe, we determined the times of bursts from a public lightcurve extracted from the event files using standard procedure. We calculated  the fluences by correlating the count fluences to the fluences for bursts also observed by \insight.
From this analysis we assembled 14 epochs of burst data (Table \ref{tab:srg_epochs}).

\begin{table*}[b!]
 \caption{Averaged burst properties of \srga\ during (approximately) daily epochs of the 2024 outburst, derived from burst measurements made by \ixpe\ and \insight. Each row lists the MJD midpoint of the burst group, and the averaged observational quantities and standard deviation: bolometric fluence, $\alpha$-value (ratio of persistent to burst flux over the burst interval), and bolometric persistent flux. The recurrence time $\Delta t_i$ and uncertainty { for each daily epoch $i$} are derived from a linear fit to the burst arrival times, and the number of bursts as well as the fit \ac{RMS} error are listed in the last two columns. The epochs marked with an asterisk "*" are those selected for the {\tt base\_sel}, 
 {\tt base\_sel\_mr} and {\tt base\_mr\_prior} runs (see Table \ref{tab:runs}).}\label{tab:srg_epochs}
 \begin{tabular}{@{\extracolsep{\fill}}lcccccc}
   \toprule
Burst epoch & Bolometric fluence & $\alpha$ & Bolometric persistent flux & $\Delta t$ & No. bursts & Fit RMS\\
(MJD) & ($10^{-6}\ {\rm erg\,cm^{-2}}$) & & ($10^{-9}\ {\rm erg\,cm^{-2}\,s^{-1}}$) & (hr) & & (hr)\\
     \hline
60363.65801* &  $0.285\pm0.020$ &  $69.85\pm4.78$  &  $3.593\pm0.171$ &  $1.579\pm0.006$ &  10 & 0.052 \\
60364.36262  &  $0.296\pm0.012$ &  $70.67\pm3.38$  &  $3.583\pm0.046$ &  $1.627\pm0.003$ &  8 & 0.022 \\
60365.32604  &  $0.295\pm0.038$ &  $64.30\pm9.30$  &  $3.030\pm0.285$ &  $1.851\pm0.013$ &  7 & 0.095 \\
60366.48186* &  $0.296\pm0.013$ &  $73.28\pm8.47$  &  $2.969\pm0.221$ &  $2.022\pm0.003$ &  8 & 0.024 \\
60367.53183  &  $0.315\pm0.027$ &  $78.82\pm8.42$  &  $3.073\pm0.195$ &  $2.137\pm0.011$ &  11 & 0.109 \\
60368.47138  &  $0.312\pm0.026$ &  $75.12\pm11.58$ &  $2.727\pm0.359$ &  $2.350\pm0.012$ &  13 & 0.101 \\
60369.43403  &  $0.296\pm0.024$ &  $65.55\pm13.30$ &  $2.111\pm0.373$ &  $2.737\pm0.013$ &  15 & 0.129 \\
60370.66803  &  $0.294\pm0.035$ &  $90.66\pm12.23$ &  $2.554\pm0.153$ &  $2.854\pm0.009$ &  5 & 0.032 \\
60371.53694* &  $0.295\pm0.026$ &  $78.01\pm16.96$ &  $2.062\pm0.431$ &  $3.183\pm0.022$ &  8 & 0.085 \\
60372.43768  &  $0.311\pm0.032$ &  $82.48\pm6.16$  &  $2.144\pm0.148$ &  $3.342\pm0.027$ &  7 & 0.122 \\
60373.41549  &  $0.304\pm0.019$ &  $88.25\pm16.98$ &  $2.061\pm0.437$ &  $3.727\pm0.052$ &  8 & 0.236 \\
60374.49261  &  $0.294\pm0.009$ &  $86.21\pm9.87$  &  $1.760\pm0.177$ &  $3.987\pm0.017$ &  3 & 0.021 \\
60375.51728* &  $0.279\pm0.021$ &  $76.92\pm8.73$  &  $1.290\pm0.283$ &  $4.779\pm0.085$ &  5 & 0.208 \\
60376.44539* &  $0.289\pm0.034$ &  $75.25\pm8.66$  &  $0.935\pm0.095$ &  $6.731\pm0.130$ &  3 & 0.106 
   \botrule
    \end{tabular}
\end{table*}

\section{Method}
\label{sec:method}

The original {\sc beansp} mode of operation is to simulate every 
burst (observed or not) within an interval spanned by all the observed events \cite[]{goodwin19c}; obviously for ``clocked'' bursting behaviour this mode is wildly inefficient, as there is typically very little burst-to-burst variation during such episodes (by definition). Using the burst ``ensembles'' described in Section \ref{sec:data} is much more efficient, as only a single model evaluation need be made for each burst epoch. 

In ``ensemble'' mode, {\sc beansp} calculates the accretion rate $\dot{m}_i$ for each epoch, given the estimated bolometric persistent flux and model parameters $d$, $\xi_p$, \mns, \rns. The accretion rate, along with the other system parameters including $X$, \zcno\ and $Q_b$, are then passed to {\sc settle} to determine the predicted burst recurrence time $\Delta t_i$, energy, and $\alpha_i$. Only one model evaluation is necessary for each epoch, and the bursts predictions are independent, in contrast to ``train'' mode. The predicted quantities are converted to observational quantities (including the effects of distance, anistropy etc.) and compared against the observations, to generate an overall likelihood comprising contributions for all epochs. The user can choose whether to include specific parameters, for example $\alpha$ (see below). The user can also apply a specific prior for individual sources depending on the degree of knowledge about the system parameters.

The present work takes advantage of a number of new developments of the code and simulation approach.
We have introduced the option to use different burst models than the default, {\sc settle}. For now, the choice is limited to the interpolated {\sc kepler} grid used by \cite{johnston20} also to match the bursts from \gs. However, the capability now exists to incorporate user-defined models.

In previous work we have gone back and forth about including the $\alpha$ values in the likelihood. For the ``train'' mode, accurate measurement of $\alpha$ also requires unambiguous knowledge of the recurrence time (i.e. no missed bursts since the previous one), and the evolution of the persistent flux over the pre-burst interval, neither of which can be routinely achieved for the low observational duty cycles and rapidly-declining persistent fluxes. For that reason, we have omitted it in the most recent study \cite[]{gal24}. A further objection is that since $\alpha$ is defined from the fluence, recurrence time, and persistent flux, and these parameters are already represented in the likelihood (persistent flux as the determinant of the accretion rate), the observed $\alpha$ is not strictly independent of the other measurables, and should not be included to avoid ``double counting'' the data.
On the other hand, $\alpha$ measured from ``clocked'' burster intervals are likely more precise, and perhaps can serve as an additional Bayesian {\it prior} that usefully constrain the walkers in the multi-dimensional parameter space. For this study we use the $\alpha$ values in some runs but also omit them to determine the effect on the \ac{MCMC} walker evolution and the resulting posteriors (see Section \ref{subsec:sgra} and Appendices \ref{sec:alphas} and \ref{sec:simulation}). 

Appendix \ref{sec:alphas} in particular revisits the analysis of the previous two sources, including improved estimates of the $\alpha$ values. In short, we find that including $\alpha$, where it can be reliably determined, may provide slightly better constraints on system parameters, but also offers a useful consistency check of the overall model solutions.

\subsection{MCMC runs}
\label{subsec:mcmc}

We used the {\sc emcee} code of \cite{emcee13} for our \ac{MCMC} runs, 
as described in Table \ref{tab:runs}. Input ({\tt .ini}) files, and the last $10^5$ samples from each of these runs are available in the online data accompanying this paper\footnote{\url{https://doi.org/10.26180/33065816}}.

We generally used the default prior function for {\sc beansp}, {\tt prior\_func}, that implements flat priors for each parameter, as listed in Table \ref{tab:srga_results}. We made a few exceptions for specific runs, for consistency with earlier analyses (see e.g. Appendix \ref{sec:alphas}). We note that the  prior upper limit on the persistent emission anisotropy factor $\xi_p=2$ for earlier versions of {\sc beansp} implicitly excludes higher inclinations 
\cite[$\cos i\lesssim0.35$, or $i\gtrsim70^\circ$, according to the modelling of][]{he16}; to accommodate the possibility of higher inclinations, we increased this limit to 10.

We  note that there is no direct link between the anisotropy parameters and the inclination; i.e. $\xi_b$ and $\xi_p$ can take any combination of values, including those formally inconsistent with any of the models of \cite{he16}. The important point here is that inclination and/or anisotropy constraints can relatively easily be incorporated into {\sc beansp} runs for other sources.

\subsection{Application to \srga}
\label{subsec:srga}

We then applied the {\sc beansp} code to the ensemble mode data assembled in Section \ref{sec:data}. We carried out several runs, varying the input data, priors,
and treatment of neutron star mass and radius, as summarised in Table \ref{tab:runs}. Each of these runs used the {\sc settle} model, as the limited {\it kepler} grid used for \gs\ likely would not cover either the range of parameters nor accretion rates required to match \srga.

The inclination for the \srga\ system has been estimated by a number of methods. Fitting the reflection spectrum led \cite{mandal25} to derive $i=(50.3^{+2.0}_{-1.3})^\circ$, consistent with the value adopted by \cite{malacaria25} based on the same approach \cite[see also][]{molkov24}. Notably, broadband spectral modelling by \cite{li25} yielded two different inclination estimates depending on the observation epoch: a poorly constrained $i=31^{+11}_{-14}{^\circ}$ during an epoch with shorter hard X-ray exposure, and a higher, better-constrained value of $i=68^\circ \pm 5^\circ$ during a second epoch.  \cite{papitto25} instead modelled the phase-resolved polarized emission measured by \ixpe, obtaining an even higher value at $i=(74.1^{+5.8}_{-6.3})^\circ$.

The possibility of a higher inclination is particularly problematic for our prior choice, as this would imply significant suppression of the observed luminosity based on interactions with the disk, also with implications for the distance. For example, based on the modelling of \cite{he16}, 
anisotropy factors of 1.21 and 1.88 
for the burst and persistent emission (respectively, for an inclination of $68^\circ$ as an example) might be expected; with the burst anisotropy factor $\xi_b$ potentially as high as $\approx2$, the implied distance based on the peak fluxes of bursts by \cite{fu25} and others may be lower by a factor of up to $\sqrt{\xi_b|_{\rm max}}\approx1.4$.
In lieu of more certainty about the inclination and distance we retain the flat priors for $\xi_b$ and $\xi_p$ for most runs, but experiment with one more restricted prior in Section \ref{sec:results}. 

We also generated a set of simulated data (using the {\tt sim\_data} method of {\sc beansp}) based loosely on the properties of \srga, and attempting to replicate the range of properties of the bursts observed during the 2024 outburst. Here the goal was to recover the parameters in the ideal case where we know that the model correctly describes the data. 
We describe the results of the simulation study in Appendix \ref{sec:simulation}.

\subsection{Cross-calibration with {\sc kepler} results}
\label{subsec:crosscal}

To test the new comparison method, we  applied it to the best-known source that has exhibited ``clocked'' bursting, \gs.
As the {\sc settle} model is not expected to give reliable results for systems with high H-fraction burning H-rich fuel (case V of \citealt{gal21a}), we  adopted instead the grid interpolation model of \cite{johnston20}. There are several differences between our treatment and the previous work.

First, we used a common mapping of (bolometric) persistent flux to accretion rate $\dot{m}$, such that the ratio of persistent fluxes between the different pairs of epochs would be identical to the ratio of accretion rates. \cite{johnston20} instead adopted a separate accretion rate for each epoch, which could vary independently in their calculation. Such an approach might be necessary if there were systematic errors in the estimation of the bolometric flux \cite[e.g.][]{thompson08}.

Second, we used a common base flux $Q_b$ for each epoch, while \cite{johnston20} adopted separate values allowed to vary independently. As they argue, this quantity is expected to vary as a function of accretion rate. On the other hand, we might expect systematic differences between the {\sc kepler} and {\sc settle} $Q_b$ values, due to the different model architectures; the two parameters may not measure the same quantity (and variability in one does not necessarily imply variability in the other). For example, a best-fit $Q_b$ for {\sc settle} may include contributions that would be more correctly modelled as steady H-burning at the base in {\sc kepler}.

Third, we include the distance $d$, and burst and persistent flux emission anisotropy ($\xi_b$ and $\xi_p$, respectively; see \citealt{concord22}) as separate parameters, while \cite{johnston20} combined these as $d\sqrt{\xi_b}$ and $\xi_p/\xi_b$. Functionally these changes likely have little effect, and we can derive the combined quantities for comparison purposes.

Fourth, we used a flat prior for \zcno, rather than the $\beta$-distribution adopted by \cite{johnston20}. The other priors were all flat with boundaries corresponding to the limits of the model grid.

Fifth, we did not include the peak burst flux in the likelihood calculation, either for the ensemble sample bursts, nor for the PRE burst observed by \cite{chenevez16}.

We also used the burst simulations in Table 2 of \cite{johnston18} to provide synthetic burst data based on the 2002 outburst of \sax\ to {\sc beansp}; these bursts were simulated with 
$X_0=0.44$, $\zcno=0.02$, $Q_b=0.3$~Mev/nucleon, $d=3.5$~kpc, $\xi_p= 1.1$, $M_{\rm NS}=1.4\ M_\odot$ and $R_{\rm NS}= 11.2$~km. To replicate the observed sample, we only included bursts B1, B4, B5 \& B6. 
No anisotropy was assumed for the burst emission, which corresponds to $\xi_b=1$.
The runs on these data were performed in ``train'' mode, listed in Table \ref{tab:runs} ({\tt johnston18c} and {\tt johnston18}), and with results described in appendix \ref{subsec:kepler}.

\section{Results}
\label{sec:results}

For each \ac{MCMC} run (see Table \ref{tab:runs}), we started with 
walker positions distributed initially within a small hypersphere in parameter space.  We ran each trial typically for $10^4$ steps initially, observing the evolution of the distribution of the walker values, and the comparison between the observations and the model-predicted values for the latest set of walker positions.
Throughout the runs we attempt to maintain the acceptance fraction $f_{\rm acc}$ in the range 0.2--0.5. Usually this fraction would start out $>0.5$ and decrease throughout the run; reducing the {\tt stretch\_a} parameter, which quantifies the scale of the ``stretch move'' algorithm for updating the walker positions \cite[]{gw10}, has the effect of stopping (or even reversing) this decrease.

{ {\sc beansp} outputs the  autocorrelation time provided by {\sc emcee}, and also calculates an integrated estimate following \cite{gw10}. However, neither of these values indicated convergence for the runs in this paper, regardless of how many steps they were run for. This issue was also noted for the previous application to \sax\ and IGR~J17498$-$2921 \cite[]{gal24}. In lieu of a quantitative convergence criterion, we continued each run until the posterior distributions reached a relatively stable level and no further significant evolution occurred. We then evaluate the posteriors excluding the initial ``burn-in'' phase of each run.}

For some runs the walkers spread out into two or more discrete concentrations in parameter space; such behaviour may indicate a multi-modal likelihood hypersurface, which { must} complicate the search for the global best fit. 
{ In many cases once the walkers separate into such groups, they remain close to those loci, 
even when a region with superior likelihood (i.e. better agreement between observations and model predictions) is already populated by other walkers. To overcome this issue, }
the {\sc beansp} code allows the user to ``partition'' the samples to visualise the posterior distributions of walker subsamples, and compare how well each group replicates the observations. In cases where one group offers much better agreement with the observations, the user can ``prune'' the walkers to follow only the best-fitting subgroup for further steps.

Below we describe the results of specific applications of the method. Our results take the form of posterior distributions of the model parameters, taken from the last few 100,000 samples of the run (i.e. 1000 steps of the 100- or 500-walker samples), and the comparison of the predicted burst parameters for models with parameters corresponding to those samples. 

\begin{table*}[b!]
 \caption{List of {\sc beansp} runs reported in this paper. ``Source'' refers to the burst source, or ``simulated'' for the various simulated datasets (see Appendix \ref{sec:simulation}). The run label serves to uniquely identify the run for that source; the numerical model used is listed in the third column ({\sc settle} is the {\sc beansp} default). The fourth and fifth columns lists the number of \ac{MCMC} walkers and the number of steps for the run; the sixth column specifies whether or not the $\alpha$ values were included in the likelihood. The seventh column describes the treatment of neutron star mass \mns\ and radius \rns; ``fixed'' indicates the values were set at the canonical values of $1.4\ M_\odot$ and $11.2$~km, respectively, while ``free'' indicates they were free to vary. Finally the last two columns list the {\tt stretch\_a} parameter used in the {\sc emcee} implementation, and the final average acceptance fraction $f_{\rm acc}$.}\label{tab:runs}
 \begin{tabular}{@{\extracolsep{\fill}}lllcccccc}
   \toprule
Source & Run label & Model & No. walkers & No.steps & $\alpha$ & \mns, \rns & {\tt stretch\_a} & $f_{\rm acc}$ \\
     \hline
\sax$^{{\rm a}}$ & {\tt base\_newalpha} & {\sc settle} & 500 & $10^4$ & Y & free &  1.125 & 0.266\\
{\sc kepler} simulated$^{{\rm b}}$ & {\tt johnston18c} & {\sc settle} & 100 & $10^4$ & N & fixed & 1.125 & 0.264 \\
                      & {\tt johnston18} & {\sc settle} & 100 & $10^4$ & N & free & 1.125 & 0.2407\\
\igr$^{{\rm a}}$ & {\tt base\_alpha} & {\sc settle} & 500 & $1.2\times10^4$ & Y & free & 1.063 & 0.212\\
\gs & {\tt grid\_mr} & {\sc kepler}$^{{\rm c}}$ & 100 & $5\times10^4$ & N/A & free & 1.063, 0.243 \\
\srga & {\tt base} & {\sc settle} & 100 & $5\times10^4$ & Y & fixed & 1.125 & 0.227\\
& {\tt base\_sel} & {\sc settle} & 100 & $10^4$ & Y & fixed & 1.25 & 0.229 \\
& {\tt base\_sel\_mr} &{\sc settle} & 500 & $2\times10^4$ & Y & free & 1.125 & 0.227\\
& {\tt base\_mr\_prior} & {\sc settle} & 100 & $10^4$ & Y & free &  1.25 & 0.228 \\
{\sc settle} simulated$^{{\rm b}}$ & {\tt sim1} & {\sc settle} & 100 & $5.2\times10^4$ & N & fixed & 1.25 & 0.242\\
          & {\tt sim1\_alpha} & {\sc settle} & 100 & $4\times10^4$ & Y & fixed & 1.25 & 0.234\\
          & {\tt sim1\_alpha\_mr} & {\sc settle} & 500 & $3\times10^4$ & Y & free &  1.125 & 0.244\\
          & {\tt sim1\_prior} & {\sc settle} & 500 & $6.2\times10^4$ & Y & free & 1.25 &  0.268
   \botrule
    \end{tabular}
    \begin{tabnote}
    {$^{{\rm a}}$ See Appendix \ref{sec:alphas}.}\tnp
    {$^{{\rm b}}$ See Appendix \ref{sec:simulation}.}\tnp
    {$^{{\rm c}}$ Interpolated, using the grid of \cite{johnston20}; see text for details.}\tnp
    \end{tabnote}
\end{table*}

\subsection{SRGA 144459.2$-$604207}
\label{subsec:sgra}

We first applied {\sc beansp} to the full set of burst ensemble data from \srga\ (Table \ref{tab:srg_epochs}). The initial run, {\tt base},
gave rise to some outliers in the walker distribution within
the first $10^4$ steps. Thus, we pruned the walker positions to retain the fraction which best-fit the observations, continuing for another $4\times10^4$ steps.

The model tended to over-predict the longer recurrence times observed { for the last two}  epochs; that is, the predicted recurrence times were longer than observed, by about 14\%. { For comparison, the average of the ratio between the predicted and observed recurrence times for the other epochs was $0.98\pm0.08$. We assessed the overall goodness of fit by calculating the \ac{RMS} error between the predicted and observed recurrence times, i.e.
\begin{equation}
{\rm RMS} = \sqrt{\sum^N\frac{(\Delta t_i-\Delta t_{i, {\rm pred}})^2}{N}}    
\end{equation}
where $\Delta t_{i, {\rm pred}}$ is the predicted recurrence time for epoch $i$, and $N$ is the number of epochs included.}
The overall \ac{RMS} error between the observed and predicted recurrence times, adopted as a rough measure of model accuracy, was 
{ 0.377~hr}. 
We attribute 
the excessive mismatch for the long-recurrence time epochs
to the presence of
more epochs with short ($<3$~hr) recurrence times in the full sample, ``driving'' the walkers towards parameter regions which preferentially match those epochs.  

To overcome this bias to short recurrence times, we downselected the epochs in an attempt to enhance the constraints from the later observations (with longer recurrence times), retaining only those epochs marked with an asterisk in Table \ref{tab:srg_epochs}. The first run with the restricted set of epochs, {\tt base\_sel}, 
substantialy reduced the overprediction factor for the longer recurrence times, and also gave a reduced \ac{RMS} error of 
{ 0.226~hr}. Thus, we 
used this data selection as the basis of subsequent runs for this source.

\begin{figure}
  \includegraphics{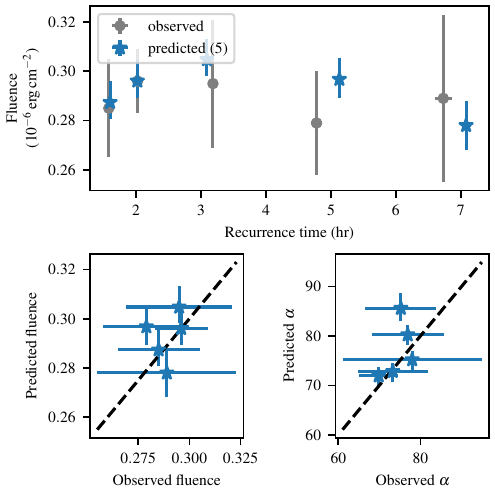}
  \caption{Comparison of observed and predicted burst properties for \srga\/ from the last $10^5$ samples of run {\tt base\_sel}. The top panel 
  { shows} 
  }
  \label{fig:base_sel_comparison}
\end{figure}

The agreement of the model predictions with the observations is quite good (Figure \ref{fig:base_sel_comparison}). The model-predicted fluences were consistently within the observed uncertainty ranges, although the recurrence times for the two later epochs were still over-predicted by 5--7\%. The predicted $\alpha$ values agreed well, although also tend to be over-predicted for the last two measurements in the selected dataset, by up to 13\%.

\begin{figure*}
  \includegraphics{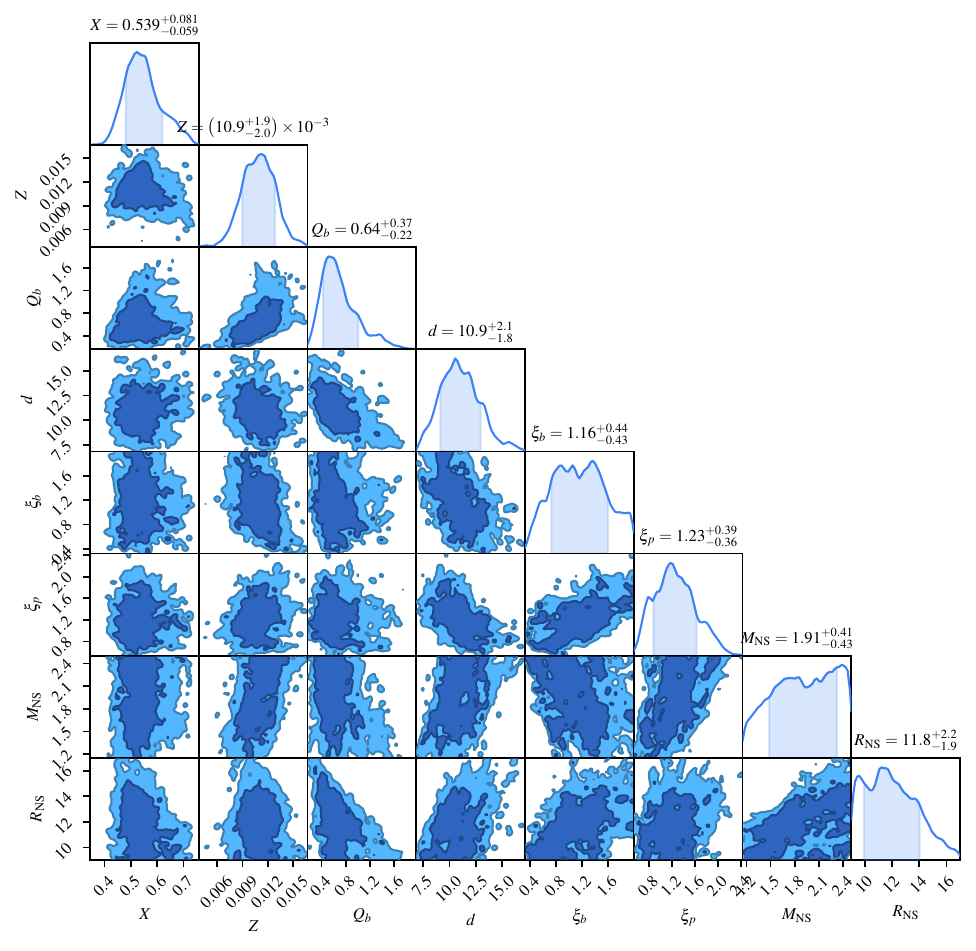}
  \caption{Posterior distributions for \srga\/ from the last $5\times10^5$ samples of run {\tt base\_sel\_mr}. Each panel shows the 2-dimensional distribution of posteriors for each model parameter taken in pairs: hydrogen mass fraction $X$, CNO metallicity $\zcno$, base flux $Q_b$, distance $d$, burst and persistent anisotropy factor $\xi_b$ and $\xi_p$, respectively; and neutron star mass \mns\ and radius \rns. Each diagonal panel shows the overall posterior distribution for each parameter alone. 1- and 2-$\sigma$ confidence regions are shown as dark and light blue, respectively.}
  \label{fig:base_sel_posteriors}
\end{figure*}

With this encouraging initial result, we freed the neutron star mass $\mns$ and radius $\rns$ in the run {\tt base\_sel\_mr}, extending an initial exploratory run to $2\times10^4$ steps with 500 walkers. We found essentially no change to the comparison of the observed and predicted burst parameters, including the \ac{RMS} error, compared to the {\tt base\_sel} run shown in Figure \ref{fig:base_sel_comparison}.
The posteriors for {\tt base\_sel\_mr} show reasonably concentrated distributions for $X$, $\zcno$, $Q_b$ and $d$, although the other parameters are less well constrained (Figure \ref{fig:base_sel_posteriors}). The corresponding best parameter estimates and $1\sigma$ errors for all the runs for \srga\ are listed in Table \ref{tab:srga_results}.

\begin{figure}
  \includegraphics{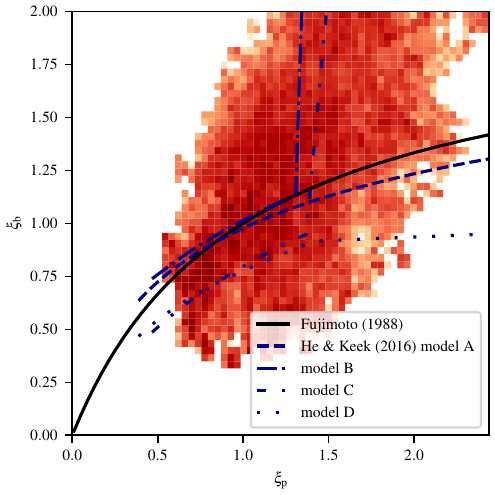}
  \caption{Combined posterior distributions for the anisotropy parameters $\xi_b$ and $\xi_p$ \srga\/ from the last $5\times10^5$ samples of run {\tt base\_sel\_mr}. We over-plot the predicted curves for various theoretical models as a function of inclination including flat and flared disks, from \cite{fuji88} and \cite{he16}. }
  \label{fig:base_sel_xis}
\end{figure}

The anisotropy parameters $\xi_b$ and $\xi_p$ were relatively poorly constrained. It is appealing to use the derived constraints to distinguish between the two possible inclination ranges 
{ established in previous studies (see Section \ref{subsec:srga})}. 
The relatively low values of both anisotropy parameters would seem to favour the lower inclination range, but the situation may not be this clear-cut. A comparison of our combined posteriors with the predictions of models by \cite{fuji88} and \cite{he16} suggest that the $\xi_b$ is typically too large compared to the $\xi_p$ value for consistency with the models (Figure \ref{fig:base_sel_xis}).

Naturally, it is also possible that those models do not accurately describe the disk anisotropy in \srga. There is no compelling evidence that the disk is truncated at the inner edge \cite[]{malacaria25,mandal25} where a gap might significantly alter the anisotropy properties, but this possibility must be considered. 

We made one final test where we also used the same prior on $\xi_b$, $\xi_p$ as for the simulated data. This run, {\tt base\_mr\_prior} was successful in the sense that the posterior distributions included the prior centroids for those parameters, $\xi_b=1.9$, $\xi_p=2.96$ (Table \ref{tab:srga_results}), and further drove the distance to significantly smaller values (a consequence of the increase in $\xi_p$); but also required a very massive and compact neutron star, with consequently excessive surface gravity and redshift. (The posterior distributions of \mns\ and \rns\ for this run were substantially limited by the upper and lower prior limits, respectively).

\begin{table*}[b!]
 \caption{Inferred system parameters for \srga. In each row we list the model parameter, units, prior used as input to the {\sc beansp} runs, and the 1-$\sigma$ confidence interval from the  posterior distributions for each run, labeled as in Table \ref{tab:runs}. The final row, labeled \ac{RMS}, quantifies the residual between the predicted recurrence time and the measurements averaged over all epochs, and is an indicative quantity for the quality of the accuracy of the model predictions. \label{tab:srga_results}}
\begin{tabular}{cccccccc}
\toprule
Parameter & Units & Prior & {\tt base } & {\tt base\_sel} & {\tt base\_sel\_mr} & {\tt base\_mr\_prior} \\
\hline
$X$ &  & $U[10^{-4},0.76]$ & $0.57\pm0.05$ & $0.54_{-0.04}^{+0.06}$ & $0.54_{-0.06}^{+0.08}$ & $0.53_{-0.05}^{+0.07}$  \\
\zcno &  & $U[10^{-4},0.056]$ & $0.0068_{-0.0003}^{+0.0005}$ & $0.0099\pm0.0014$ & $0.0109\pm0.0019$ & $0.015\pm0.002$ \\
$Q_{\rm b}$ & MeV/nucleon & $U[10^{-6},5]$ & $0.67_{-0.06}^{+0.07}$ & $0.8_{-0.2}^{+0.3}$ &  $0.6_{-0.2}^{+0.4}$ & $0.67_{-0.15}^{+0.26}$  \\
$d$ & kpc & $U[1,20]$ & $9.9_{-0.7}^{+1.2}$ & $10.5_{-1.0}^{+1.4}$ & $10.9_{-1.7}^{+2.1}$ & $6.4\pm0.5$ \\
$\xi_b$ & & $U[0.01,2]^{{\rm a}}$ & $1.5_{-0.3}^{+0.4}$ & $1.2\pm0.3$ & $1.2\pm0.4$ & $1.36_{-0.19}^{+0.31}$ \\
$\xi_p$ & & $U[0.01,10]^{{\rm a}}$ & $1.10_{-0.24}^{+0.17}$ & $0.9\pm0.2$ & $1.2\pm0.4$ & $3.6_{-0.7}^{+0.6}$ \\
\mns & $M_\odot$ & $U[1.15,2.5]$ & (fixed) & (fixed) & $1.9\pm0.4$ & $2.34_{-0.29}^{+0.12}$ \\
\rns  & km & $U[9,17]$ & (fixed) & (fixed) & $11.8_{-1.9}^{+2.2}$ & $9.4_{-0.3}^{+0.6}$\\
$g^{{\rm b}}$ & $10^{14}\ {\rm cm\,s^{-2}}$ & & & & $2.3_{-0.6}^{+1.4}$ & $6.4\pm1.5$ \\
$1+z^{{\rm b}}$ & & & & & $1.36_{-0.08}^{+0.18}$ & $1.9\pm0.2$  \\
\hline
RMS & hr & & { 0.377} & { 0.226} & { 0.231} & { 0.216}
\botrule
    \end{tabular}
    \begin{tabnote}
    {$^{{\rm a}}$ Excludes the {\tt base\_mr\_prior} run, which used a specific prior on each of $\xi_b$, $\xi_p$, as described in the text}\tnp
    {$^{{\rm b}}$ Derived parameter from the combined distribution of \mns, \rns\ (where free to vary).}\tnp
    \end{tabnote}
\end{table*}

\subsection{GS 1826$-$24}
\label{subsec:gs1826}

We performed a single run ({\tt grid\_mr} in Table \ref{tab:runs}) with 100 walkers, for 50,000 steps in total. We did not compare the $\alpha$ values, as these are not provided by the {\sc kepler} grid interpolation. After the initial 10,000 steps the presence of multiple loci in $X$--$\zcno$ space led us to prune the walkers, retaining only the highest $X$-values. 

\begin{figure}
    \includegraphics[width=9cm]{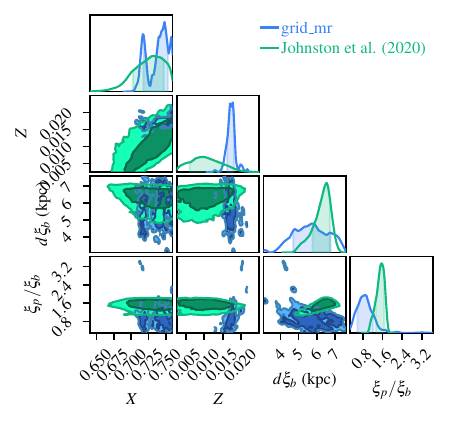}
    \caption{Posterior distributions (blue shaded contours) for selected parameters for GS~1826$-$24, derived from the comparison of the {\sc kepler} grid interpolation model to the 3-epoch reference data. We consider here only the common parameters of interest, $X$, $Z$ and derived parameters for the {\sc beansp} implementation, $d\xi_b$ and $\xi_p/\xi_b$. The corresponding posteriors derived by \cite{johnston20} are shown as green shaded regions and contours.}
    \label{fig:grid_mr}
  \end{figure}

We compare the derived posteriors with those of \cite{johnston20}  in Figure \ref{fig:grid_mr}. There is reasonably good overlap between the posteriors, although the {\sc beansp} runs sometimes cover a smaller or larger region. Multiple peaks present in the $X$-posterior appear to fall in-between the values at which the grid is evaluated, i.e. $X= [\ldots, 0.7, 0.73, 0.76]$, which may indicate an issue with the interpolation scheme.

Given the differences in treatment to that of \cite{johnston20}, we consider the {\sc beansp} implementation broadly successful.

\section{Discussion}
\label{sec:disc}

We have demonstrated a modified approach for comparing thermonuclear burst observations to models, using a range of burst models. The method takes advantage of regular, consistent (``clocked'') bursting and attempts to replicate the average properties of bursts within one or more observation epochs: recurrence time, fluence, and (optionally) $\alpha$-value, given the measured persistent flux at each epoch.

We applied the new method to 14 daily epochs of data from the profligate burster \srga, assembled from a sample of 94 unique events observed by \insight\ and \ixpe\ during the 2024 outburst. We performed most of our runs with a subsample of five epochs to encourage better agreement with the longer recurrence times measured late in the outburst. We obtained reasonably good agreement with the observations, with the model slightly overestimating the burst recurrence time for the two { epochs with the lowest accretion rates } in the restricted sample, . 

The posterior distributions for all the runs exhibited slightly sub-solar H-fraction $X$ and CNO metallicity $\zcno$, consistently $X\simeq0.54$ and $\zcno\simeq0.01$. 
The H-fraction value is similar to that inferred for \sax\ (Appendix \ref{sec:alphas}) and may arise from similar evolution of the mass donor in \srga\ \cite[]{goodwin20a}.
{ In contrast to \sax, } the combination of inferred H-fraction and frequent bursts with $\Delta t \gtrsim1.6$~hr places the ignition conditions for \srga\ firmly in case V of \cite{gal21a}, i.e. He-ignition in a H-rich environment. The estimated time to exhaust H at the base of the fuel layer through steady burning for the parameters inferred for \srga\ is roughly
\begin{equation}
    t_{\rm CNO}=9.8\left(\frac{X}{0.7}\right)\left(\frac{\zcno}{0.02}\right)^{-1}=14_{-2}^{+4}\ {\rm hr}
\end{equation}
\cite[e.g.][]{concord22}, substantially longer than the recurrence times in our sample, and ensuring that plenty of hydrogen is still present at the base of the fuel layer at the point of ignition.

{ We point out two principal reasons to view the inferred system parameters with caution. First, the lack of (quantitative) convergence for the MCMC runs (see Section \ref{sec:results}) leaves some doubt about the reliability of the posterior distributions, even though we continued each run for long enough to establish a {\it qualitative} equilibrium in the walker positions. This issue has arisen for each of the applications of the updated version of {\sc beansp}, as described in \cite{gal24}. We plan future investigations to try to resolve this issue, but we surmise that there may be some idiosyncrasy with the algorithm that prevents it reaching conventional convergence criteria.

Second, as {\sc settle} is understood to poorly replicate ignition in a H-rich environment \cite[e.g.][]{gal06c}, the posterior distributions may be subject to systematic errors due to model mismatch. }
Notably the inferred H-fraction was also
in excess of the 0.4 limit derived by \cite{fu25}, which may indicate an upward bias on this parameter due to { these limitations of the model}. 
We speculate that similar burst properties could be achieved with lower $X$ values (i.e. consistent with this limit), in a model which fully takes into account the heating effect of steady burning prior to ignition, also providing ``thermal inertia'' which allows earlier ignition \cite[e.g.][]{heger07b}.
We suggest that the { larger} model mismatch for the longer recurrence time epochs may also arise from { {\sc settle}'s limitations for H-rich burst ignition predictions

Other than we discuss above, the derived parameter constraints are broadly compatible with other published work.
\cite{dohi25} also concluded non-solar elemental abundances in \srga\ based on comparison of the observed burst lightcurves with the predictions of a general relativistic stellar evolution code, {\sc heres}. They inferred $X/Y\approx1.5$, where $Y$ is the fuel He-fraction (i.e. $1-X-\zcno$; solar composition would have $X/Y>2$), which is fully compatible with the posterior distributions from {\sc settle}. On the other hand, their preferred \zcno\ value was in excess of solar, but produced longer recurrence times than observed, for standard rates for certain key nuclear reactions.

\cite{dorsman26} undertook pulse-profile modelling of the source, a completely independent way of constraining the properties but with common parameters of mass, (equatorial) radius, distance and inclination. The analysis was complicated by a large number of additional model parameters describing the geometry of the pulsing region, as well as several options for data treatment and impact of prior bounds. Inferred masses tended to be around $2\ M_\odot$ or above, and with radii in the range 11--14~km. However their distance posteriors were consistently centred around 3~kpc, substantially lower than the values inferred here. This result was despite a normal distance prior centred on 8.5~kpc; as those authors point out, this discrepancy may be attributable to a missing component in their radiation model, perhaps related to disk reflection.}

The inclusion of a prior { in the {\sc settle} comparison,} intended to reflect one of the inclination constraints (derived from polarisation measurements made by \citealt{papitto25}), results in a lower distance but unrealistically massive and compact neutron star. 
It is clear that constraints on the distance, which is often one of the most sought-after quantities for such systems, depend inextricably on assumptions about the degree of anisotropy of the emission.

{ This study has also resolved several questions regarding the details of the {\sc beansp} implementation.}
Inclusion of $\alpha$ in the likelihood when comparing models to observations (in parallel with previously-analysed sources with sub-solar $X$ values) confirms that this parameter, when measured reliably, provides an important consistency check for model fidelity. 
We tested the method on simulated data, demonstrating that the approach can replicate the input parameters, irrespective of the inclusion (or not) of the $\alpha$ parameters; however, the example chosen led to only weak constraints on the neutron star parameters, including the burst and persistent emission anisotropy and the neutron star mass and radius. We suggest that that other ensembles, if observed over a wider range of accretion rates (and hence burst ignition regimes) might allow more stringent constraints on system properties.

We also tested the method on a ``reference'' sample from the best-known source exhibiting ``clocked'' bursting, GS~1826$-$24, comparing instead with interpolated results over a grid of pre-computed {\sc kepler} models, partially replicating the previous work of \cite{johnston20}. We achieved moderately good agreement with the posterior distributions from that study, given the differences in our approach. One significant difference is the assumed consistency of the ratio between the persistent flux measurements and the accretion rate; this restriction could easily be relaxed for future simulations with {\sc beansp}, to better match the previous approach. A more pressing question for future work is how well bolometric corrections can be estimated in ensembles that span much larger time ranges (and hence potentially different source persistent flux states) than for the \srga\ data (cf. with \citealt{thompson05}).

{ The neutron star mass \mns\ and radius \rns\ are perhaps the most sought-after parameters in these and other studies \cite[e.g.][]{ozel16b}. Unfortunately the burst ignition conditions are only weakly sensitive to the surface gravity, and the specific sensitivity of the {\sc beansp} algorithm to mass and radius is likely to be poor in general. This expectation is borne out by the more-or-less unconstrained posteriors derived from both the simulation data and the applications to \srga\ and other sources. There is some evidence for a bias to higher masses in the simulation study, but the posterior range is so large this bias is not significant. }

In future we plan to compare the \srga\ sample with a more accurate model such as {\sc kepler}, that fully treats the steady burning between bursts, and which may provide more reliable constraints on the fuel composition. This comparison is not currently possible as the existing model grids (used by \citealt{johnston20} for application to GS~1826$-$24) only cover a narrow range of $X$, excluding the parameter space where \srga\ likely falls. We  also cannot call {\sc kepler} from the code (as we do with {\sc settle}), due to the prohibitive compute time for individual runs. However, the {\sc beansp} code provides plenty of flexibility to include interpolation on more suitable model grids or incorporation of different models altogether.

\begin{acknowledgements}
We acknowledge the contributions of the anonymous referee whose careful reading and attention to detail in their review, substantially improved this paper.
This research was supported by the International Space Science Institute (ISSI) in Bern and the International Space Science Institute–Beijing (ISSI-BJ), through the joint ISSI/ISSI-BJ International Team project `Thermonuclear X-ray Bursts: From Simulations to Multi-Wavelength Observations' led by Z. Li and D. Galloway (ISSI Team project \#25-647). Z.S.L. was supported by National Natural Science Foundation of China (No. 12273030).
\end{acknowledgements}

\section*{Data Availability}

The \srga\ burst data is available on request. The ensemble average values (Table \ref{tab:srg_epochs}), and the burst data used for analysis of \sax\ and \igr\, are all available as part of the {\sc beansp} Github repository\footnote{\url{https://github.com/adellej/beans}}.
A snapshot of the code and the posterior samples from the last $10^3$ steps of each of the runs presented here can be found at the 
Monash Bridges repository\footnote{\url{https://doi.org/10.26180/33065816}}.

\bibliography{all}

\begin{thebibliography}{}

\bibitem[{Chenevez} et~al., 2016]{chenevez16}
{Chenevez}, J., {Galloway}, D.~K., {in 't Zand}, J.~J.~M., {Tomsick}, J.~A.,
  {Barret}, D., {Chakrabarty}, D., {F{\"u}rst}, F., {Boggs}, S.~E.,
  {Christensen}, F.~E., {Craig}, W.~W., {Hailey}, C.~J., {Harrison}, F.~A.,
  {Romano}, P., {Stern}, D., \& {Zhang}, W.~W. 2016, {A Soft X-Ray Spectral
  Episode for the Clocked Burster, GS 1826-24 as Measured by Swift and NuStar}.
\newblock {\em \apj}, 818, 135.

\bibitem[{Di Salvo} and {Sanna}, 2022]{disalvo22}
{Di Salvo}, T. \& {Sanna}, A.
\newblock {Accretion Powered X-ray Millisecond Pulsars}.
\newblock In {Bhattacharyya}, S., {Papitto}, A., \& {Bhattacharya}, D.,
  editors, {\em Astrophysics and Space Science Library} 2022,, volume 465 of
  {\em Astrophysics and Space Science Library}, pp. 87--124.

\bibitem[{Dohi} et~al., 2025]{dohi25}
{Dohi}, A., {Nishimura}, N., {Hirai}, R., {Takeda}, T., {Iwakiri}, W.,
  {Tamagawa}, T., {Aoyama}, A., {Enoto}, T., {Iwata}, S., {Kato}, Y.,
  {Kitaguchi}, T., {Mihara}, T., {Ota}, N., {Takahashi}, T., {Watanabe}, S., \&
  {Yamasaki}, K. 2025, {Evidence of non-solar elemental composition in the
  clocked X-ray burster SRGA J144459.2-604207}.
\newblock {\em \pasj}, 77(3), L17--L23.

\bibitem[{Dorsman} et~al., 2026]{dorsman26}
{Dorsman}, B., {Salmi}, T., {Watts}, A.~L., {Ng}, M., {Bobrikova}, A., {Marco},
  A.~D., {Galloway}, D.~K., {Guillot}, S., {Hoogkamer}, M., {Kini}, Y.,
  {Monaca}, F.~L., {Loktev}, V., {Lucchini}, M., {Malacaria}, C., {Mao}, Y.-H.,
  {Papitto}, A., \& {Poutanen}, J. 2026, {Pulse profile modelling of the 2024
  outburst of the accreting millisecond pulsar SRGA
  J144459.2{\ensuremath{-}}604207}.
\newblock {\em \mnras}, 551(1), stag1340.

\bibitem[Foreman-Mackey et~al., 2013]{emcee13}
Foreman-Mackey, D., Hogg, D.~W., Lang, D., \& Goodman, J. 2013, emcee: The mcmc
  hammer.
\newblock {\em Publications of the Astronomical Society of the Pacific},
  125(925), pp. 306--312.

\bibitem[{Fu} et~al., 2025]{fu25}
{Fu}, T., {Li}, Z., {Pan}, Y., {Ji}, L., {Chen}, Y., {Kuiper}, L., {Galloway},
  D.~K., {Falanga}, M., {Xu}, R., {Li}, X., {Ge}, M., {Song}, L.~M., {Zhang},
  S., \& {Zhang}, S.-N. 2025, {A Comprehensive Study of Type I (Thermonuclear)
  Bursts in the New Transient SRGA J144459.2{\textendash}604207}.
\newblock {\em \apj}, 980(2), 161.

\bibitem[{Fujimoto}, 1988]{fuji88}
{Fujimoto}, M.~Y. 1988, {Angular distribution of radiation from low-mass X-ray
  binaries}.
\newblock {\em \apj}, 324, 995--1000.

\bibitem[{Galloway} and {Cumming}, 2006]{gal06c}
{Galloway}, D.~K. \& {Cumming}, A. 2006, {Helium-rich Thermonuclear Bursts and
  the Distance to the Accretion-powered Millisecond Pulsar SAX J1808.4-3658}.
\newblock {\em \apj}, 652, 559--568.

\bibitem[{Galloway} et~al., 2004]{gal03d}
{Galloway}, D.~K., {Cumming}, A., {Kuulkers}, E., {Bildsten}, L.,
  {Chakrabarty}, D., \& {Rothschild}, R.~E. 2004, Periodic thermonuclear x-ray
  bursts from gs 1826-24 and the fuel composition as a function of accretion
  rate.
\newblock {\em \apj}, 601, 466--473.

\bibitem[{Galloway} et~al., 2024]{gal24}
{Galloway}, D.~K., {Goodwin}, A.~J., {Hilder}, T., {Waterson}, L., \&
  {Cup{\'a}k}, M. 2024, {Inferring system parameters from the bursts of the
  accretion-powered pulsar IGR J17498-2921}.
\newblock {\em \mnras}, 535(1), 647--656.

\bibitem[{Galloway} et~al., 2017]{gal17a}
{Galloway}, D.~K., {Goodwin}, A.~J., \& {Keek}, L. 2017, {Thermonuclear Burst
  Observations for Model Comparisons: A Reference Sample}.
\newblock {\em \pasa}, 34, e019.

\bibitem[{Galloway} et~al., 2020]{minbar}
{Galloway}, D.~K., {in't Zand}, J., {Chenevez}, J., {W{\"o}rpel}, H., {Keek},
  L., {Ootes}, L., {Watts}, A.~L., {Gisler}, L., {Sanchez-Fernandez}, C., \&
  {Kuulkers}, E. 2020, {The Multi-INstrument Burst ARchive (MINBAR)}.
\newblock {\em \apjs}, 249(2), 32.

\bibitem[{Galloway} et~al., 2022]{concord22}
{Galloway}, D.~K., {Johnston}, Z., {Goodwin}, A., \& {He}, C.-C. 2022, {Robust
  Inference of Neutron-star Parameters from Thermonuclear Burst Observations}.
\newblock {\em \apjs}, 263(2), 30.

\bibitem[{Galloway} and {Keek}, 2021]{gal21a}
{Galloway}, D.~K. \& {Keek}, L. 2021,.
\newblock {\em {Thermonuclear X-ray Bursts}}, volume 461, pp. 209--262.

\bibitem[{Goodman} and {Weare}, 2010]{gw10}
{Goodman}, J. \& {Weare}, J. 2010, {\em CAMS}, 5(3), 65--80.

\bibitem[{Goodwin} et~al., 2019]{goodwin19c}
{Goodwin}, A.~J., {Galloway}, D.~K., {Heger}, A., {Cumming}, A., \& {Johnston},
  Z. 2019, {A Bayesian approach to matching thermonuclear X-ray burst
  observations with models}.
\newblock {\em \mnras}, 490(2), 2228--2240.

\bibitem[{Goodwin} and {Woods}, 2020]{goodwin20a}
{Goodwin}, A.~J. \& {Woods}, T.~E. 2020, {The binary evolution of SAX
  J1808.4-3658: implications of an evolved donor star}.
\newblock {\em \mnras}, 495(1), 796--805.

\bibitem[{He} and {Keek}, 2016]{he16}
{He}, C.-C. \& {Keek}, L. 2016, {Anisotropy of X-Ray Bursts from Neutron Stars
  with Concave Accretion Disks}.
\newblock {\em \apj}, 819, 47.

\bibitem[{Heger} et~al., 2007]{heger07b}
{Heger}, A., {Cumming}, A., {Galloway}, D.~K., \& {Woosley}, S.~E. 2007,
  {Models of Type I X-Ray Bursts from GS 1826-24: A Probe of rp-Process
  Hydrogen Burning}.
\newblock {\em \apjl}, 671, L141--L144.

\bibitem[{Johnston} et~al., 2018]{johnston18}
{Johnston}, Z., {Heger}, A., \& {Galloway}, D.~K. 2018, {Simulating X-ray
  bursts during a transient accretion event}.
\newblock {\em \mnras}, 477, 2112--2118.

\bibitem[{Johnston} et~al., 2020]{johnston20}
{Johnston}, Z., {Heger}, A., \& {Galloway}, D.~K. 2020, {Multi-epoch X-ray
  burst modelling: MCMC with large grids of 1D simulations}.
\newblock {\em \mnras}, 494(3), 4576--4589.

\bibitem[{Li} et~al., 2025]{li25}
{Li}, Z., {Kuiper}, L., {Pan}, Y., {Xu}, R., {Chen}, Y., {Ge}, M., {Huang}, Y.,
  {Jia}, S., {Li}, X., {Song}, L., {Qu}, J., {Zhang}, S., {Tao}, L., {Feng},
  H., {Zhang}, S.-N., \& {Falanga}, M. 2025, {Timing and Spectral Studies of
  SRGA J144459.2{\ensuremath{-}}604207 with NICER, Einstein Probe, IXPE,
  NuSTAR, Insight-HXMT, and INTEGRAL During its 2024 Outburst}.
\newblock {\em \apj}, 990(1), 15.

\bibitem[{Malacaria} et~al., 2025]{malacaria25}
{Malacaria}, C., {Papitto}, A., {Campana}, S., {Di Marco}, A., {Di Salvo}, T.,
  {Cristina Baglio}, M., {Illiano}, G., {La Placa}, R., {Miraval Zanon}, A.,
  {Pilia}, M., {Poutanen}, J., {Salmi}, T., {Sanna}, A., \& {Mandal}, M. 2025,
  {Disk reflection and energetics from the accreting millisecond pulsar SRGA
  J144459.2‑604207}.
\newblock {\em \aap}, 699, A288.

\bibitem[{Mandal} et~al., 2025]{mandal25}
{Mandal}, M., {Naik}, S., \& {Chhotaray}, B. 2025, {Relativistic X-ray
  reflection and thermonuclear burst from accreting millisecond X-ray pulsar
  SRGA J144459.2-604207}.
\newblock {\em \mnras}, 542(1), 443--455.

\bibitem[{Molkov} et~al., 2024]{molkov24}
{Molkov}, S.~V., {Lutovinov}, A.~A., {Tsygankov}, S.~S., {Suleimanov}, V.~F.,
  {Poutanen}, J., {Lapshov}, I.~Y., {Mereminskiy}, I.~A., {Semena}, A.~N.,
  {Arefiev}, V.~A., \& {Tkachenko}, A.~Y. 2024, {Discovery of SRGA
  J144459.2‑604207 with the SRG/ART-XC telescope: A well-tempered bursting
  accreting millisecond X-ray pulsar}.
\newblock {\em \aap}, 690, A353.

\bibitem[{Ng} et~al., 2024]{ng24a}
{Ng}, M., {Sanna}, A., {Strohmayer}, T.~E., {Arzoumanian}, Z., {Gendreau},
  K.~C., {Ray}, P.~S., {Coley}, J.~B., {Chakrabarty}, D., {Guillot}, S.,
  {Bogdanov}, S., {Altamirano}, D., {Chenevez}, J., {Hare}, J., {Wolff}, M.~T.,
  {Guver}, T., {Jaisawal}, G.~K., {Wadiasingh}, Z., \& {Ferrara}, E.~C. 2024,
  {NICER Discovers Millisecond Pulsations and a Type I X-ray Burst from SRGA
  J144459.2-604207}.
\newblock {\em The Astronomer's Telegram}, 16474, 1.

\bibitem[{{\"O}zel} and {Freire}, 2016]{ozel16b}
{{\"O}zel}, F. \& {Freire}, P. 2016, {Masses, Radii, and the Equation of State
  of Neutron Stars}.
\newblock {\em \araa}, 54, 401--440.

\bibitem[{Papitto} et~al., 2025]{papitto25}
{Papitto}, A., {Di Marco}, A., {Poutanen}, J., {Salmi}, T., {Illiano}, G., {La
  Monaca}, F., {Ambrosino}, F., {Bobrikova}, A., {Baglio}, M.~C., {Ballocco},
  C., {Burderi}, L., {Campana}, S., {Coti Zelati}, F., {Di Salvo}, T., {La
  Placa}, R., {Loktev}, V., {Long}, S., {Malacaria}, C., {Miraval Zanon}, A.,
  {Ng}, M., {Pilia}, M., {Sanna}, A., {Stella}, L., {Strohmayer}, T., \&
  {Zane}, S. 2025, {Discovery of polarized X-ray emission from the accreting
  millisecond pulsar SRGA J144459.2{\textendash}604207}.
\newblock {\em \aap}, 694, A37.

\bibitem[{Steiner} et~al., 2018]{steiner18}
{Steiner}, A.~W., {Heinke}, C.~O., {Bogdanov}, S., {Li}, C.~K., {Ho}, W.~C.~G.,
  {Bahramian}, A., \& {Han}, S. 2018, {Constraining the mass and radius of
  neutron stars in globular clusters}.
\newblock {\em \mnras}, 476, 421--435.

\bibitem[{Thompson} et~al., 2008]{thompson08}
{Thompson}, T.~W.~J., {Galloway}, D.~K., {Rothschild}, R.~E., \& {Homer}, L.
  2008, {Deviations from the Flux-Recurrence Time Relationship in GS 1826-238:
  Potential Transient Spectral Changes}.
\newblock {\em \apj}, 681, 506--514.

\bibitem[{Thompson} et~al., 2005]{thompson05}
{Thompson}, T.~W.~J., {Rothschild}, R.~E., {Tomsick}, J.~A., \& {Marshall},
  H.~L. 2005, {Chandra and RXTE Spectra of the Burster GS 1826-238}.
\newblock {\em \apj}, 634, 1261--1271.

\end{thebibliography}

\appendix

\renewcommand{\thefigure}{\Alph{section}\arabic{figure}}
\renewcommand{\thetable}{\Alph{section}\arabic{table}}

\section{The effect of including $\alpha$}
\label{sec:alphas}

In this section we describe some experiments intended to illustrate the effect of including the observed $\alpha$ values (the ratio of the persistent to burst luminosity) in the burst model-observation comparison. In the previous study, we neglected $\alpha$ from the {\sc beansp} likelihood due to concerns about the independence of those values, coupled with the difficulty in measuring the average persistent flux over a burst interval that may include uncertainties introduced by missed bursts \cite[]{gal24}.

To partially address these issues, we introduced a new capability in the {\tt burst\_table} method which will calculate the $\alpha$-value for each observed event, based on the observed fluence, the model for the persistent flux (including any interpolation, beit linear or cubic spline as user choice) and the inferred recurrence time (including any missed bursts). The code estimates the recurrence time by dividing the interval between two observed events by $n+1$, where $n$ is the estimated (or user-defined) number of missed bursts. For \sax, these new $\alpha$-estimates were between 3 and 15\% larger than the MINBAR values.

We then re-ran the sampler on both \sax\ and \igr\ replicating the analysis of \cite{gal24}, but this time including the new $\alpha$ estimates in the likelihood. For \sax, we ran with 500 walkers and 10000 steps. We identified a bug in the code where the bolometric correction was applied in two separate locations in the conversion from persistent flux to accretion rate. This error primarily affected the distance, with the other system parameters largely consistent with the previous study (Table \ref{tab:alpha_results}). The best fit distance is now $3.2\pm0.3$~kpc, once again consistent with the value derived by \cite{goodwin19c}. 

\begin{figure}
  \includegraphics{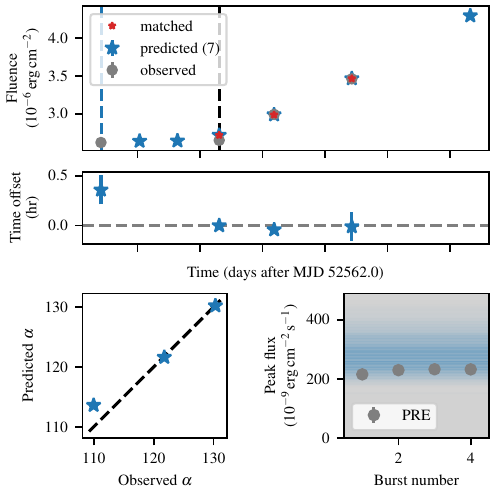}
 \caption{Predicted and observed burst properties for the last 1000 samples of run {\tt base\_newalpha} for the 2002 outburst of \sax\/ including the newly-calculated $\alpha$ values; cf. with Figure 2 of \cite{gal24}. The top panel shows the comparison between the burst arrival times and fluences; the observed bursts are shown as gray symbols, and the predictions as blue stars. No fluence prediction is made for the first burst, so the time is indicated by the vertical dashed blue line. The 3rd and 4th observed bursts are obscured by the predicted values; the reference burst is marked by the vertical black dashed line. The red-starred bursts are the predicted events chosen to match the observed; note the two additional predicted bursts observed between the first and second observed burst, and the final predicted burst, all of which fell within data gaps, and hence were not observed.
 The lower left panel shows the predicted and measured $\alpha$ values, latter incorporating our new approach via integrating the persistent flux model. 
 The lower right panel shows the peak flux values, and the blue bands show the model evaluations of the Eddington flux (assuming isotropy) at the range of distances from the last 1000 samples. The gray background indicates that this quantity was not included in the likelihood calculation.}
  \label{fig:sax_comparison}
\end{figure}

The comparison of the observed and predicted properties shows good agreement (Figure \ref{fig:sax_comparison}). The \ac{RMS} error on the observed-predicted times was 0.18~hr, an order of magnitude larger than the 1.37~min reported previously, but likely representing a compromise required to also match the $\alpha$ values. The agreement for that quantity is also good, with the notable exception of the second burst (with the lowest $\alpha$); but the recurrence time for that event was estimated as $\Delta t_{1,2}/3=15.13$~hr where $\Delta t_{1,2}$ is the time separation of the first two observed bursts. The model predicted recurrence time is significantly longer at $16.02\pm0.04$~hr, likely because the persistent flux is dropping significantly over this period.
The posterior distributionss of the parameters compare well with the previous study, 
with the notable exception of the distance, as discussed above.

For \igr, the persistent fluxes adopted were already bolometric estimates, so the bug identfied above did not affect the best-fit system parameters. The new $\alpha$ measurements exhibited a much lower variance than previously, with a mean (excluding the first and last bursts; see below) of $248\pm13$. The estimated $\alpha$ for burst \#8 was the highest but also the least precise, at $400\pm80$; this burst was observed with {\it Swift}, and there is some indication that these fluences may be significantly lower than those values measured by {\it RXTE}\/ or {\it INTEGRAL}. We re-ran the analysis as in \cite{gal24} with 12,000 steps and 500 walkers. The comparison between the observed and predicted times had an \ac{RMS} error of 0.898~hr, comparable to the 0.882~hr achieved without comparing $\alpha$. The predicted $\alpha$ well matched the newly measured values, excluding the last burst, adding weight to the suggestion that the fluence for that event was substantially underestimated.

We consider these new results to be superior to those previously published, and the current best estimates of system parameters for those objects, based on the {\sc beansp} approach. We further recommend that best estimates of observational $\alpha$ values, where available, be included in likelihoods for use with {\sc beansp} runs, to provide an important consistency check for the model.

\begin{table*}[b!]
 \caption{Inferred system parameters for \sax\ and \igr, for runs including  $\alpha$ in the likelihood. The prior on $Z$ label ``beta`` indicates the same Beta-function as adopted by \cite{goodwin19c} and \cite{gal24}. The \mns, \rns\ prior label ``steiner'' indicates the same 2-dimensional prior derived from the analysis of \cite{steiner18} for neutron stars in globular clusters, as also used by \cite{goodwin19c} and \cite{gal24}. Other details as for Table \ref{tab:srga_results}.  \label{tab:alpha_results}}
\begin{tabular}{ccccc}
\toprule
 &  &  & \sax & \igr \\
Parameter & Units & Prior & {\tt base\_newalpha} & {\tt base\_alpha} \\
\hline
\ensuremath{X} & & $U[10^{-5},0.76]$ & $0.41_{-0.10}^{+0.13}$ & $0.172_{-0.062}^{+0.156}$ \\
\zcno & & beta & $0.0151_{-0.0067}^{+0.0119}$ & $0.0035_{-0.0022}^{+0.0185}$ \\
$Q_{\rm b}$ & MeV/nucleon & $U[10^{-6},5.0]$ & $0.6_{-0.2}^{+0.3}$ & $2.6_{-0.5}^{+0.9}$ \\
$d$ & kpc & $U[1,20]$& $3.2\pm0.3$ & $4.9\pm0.4$ \\
$\xi_b$ & & $U[0.01,2]$ & $1.03_{-0.17}^{+0.20}$ & $1.83_{-0.19}^{+0.12}$ \\
$\xi_p$ & & $U[0.01,10]$ & $1.6\pm0.4$ & $1.6_{-0.3}^{+0.2}$ \\
\mns & $M_\odot$ & steiner & $1.8_{-0.4}^{+0.5}$ & $2.1_{-0.5}^{+0.3}$ \\
\rns & km & steiner & $12.2\pm1.2$ & $12.14_{-0.81}^{+1.08}$ \\
$g$ & $10^{14}\ {\rm cm\,s^{-2}}$ & & $2.2_{-0.4}^{+0.7}$ & $2.5_{-0.6}^{+0.8}$ \\
$1+z$ & & & $1.338_{-0.082}^{+0.111}$ & $1.42\pm0.11$ \\
\hline
RMS & hr & & 0.180 & 0.898 \\
\hline
\end{tabular}
\end{table*}

\section{Simulated data}
\label{sec:simulation}

Here we describe the results from our simulation study. A key stumbling block for the reliability of the {\sc beansp} burst matching algorithm is the lack of well characterised calibration data to test on. We seek to establish whether, in the case where we know the system parameters giving rise to a set of burst data, that we can recover them accurately via {\sc beansp}. A secondary objective is to understand the systematic uncertainties that may arise between different numerical models. 

To achieve these goals, we performed two separate simulation studies, the first based on ensemble mode data generated with {\sc settle} mimicking the \srga\ sample, and the second based on {\sc kepler} simulations of bursts from \sax\ intended to replicate the events observed during the 2002 outburst, presented by \cite{johnston18}. 

\subsection{\srga\ ensemble-mode}

We generated synthetic burst data with {\sc settle} based on the persistent flux values measured for \srga\ in the restricted epoch sample from Table \ref{tab:srg_epochs}, as listed in Table \ref{tab:sim_epochs}. The persistent flux values and uncertainties on all parameters are copied from the corresponding epochs of Table \ref{tab:sim_epochs}, to represent realistic observational errors. 
Note that it was not possible to fully replicate the properties of the bursts in \srga, but we attempted to come up with similar ranges of $\Delta t$ and comparable fluences and $\alpha$.

\begin{table*}[b!]
 \caption{Simulated data generated for testing {\sc beansp} in ensemble mode.}\label{tab:sim_epochs}
 \begin{tabular}{@{\extracolsep{\fill}}lcccc}
   \toprule
Burst epoch & Bolometric fluence &  & Bolometric persistent flux & $\Delta t$ \\
(MJD) & ($10^{-6}\ {\rm erg\,cm^{-2}}$) & $\alpha$ & ($10^{-9}\ {\rm erg\,cm^{-2}\,s^{-1}}$) & (hr) \\
     \hline
60363.65801  &   $0.196 \pm 0.020$ & $65.9 \pm 4.8$ & $3.593 \pm 0.171$ & $0.997 \pm 0.006$ \\
60366.48186  &   $0.195 \pm 0.013$ & $67.3 \pm 8.5$ & $2.969 \pm 0.221$ & $1.232 \pm 0.003$ \\
60371.53694  &   $0.191 \pm 0.026$ & $71.2 \pm 17.0$ & $2.062 \pm 0.431$ & $1.830 \pm 0.022$ \\
60375.51728  &   $0.182 \pm 0.021$ & $81.8 \pm 8.7$ & $1.290 \pm 0.283$ & $3.209 \pm 0.085$ \\
60376.44539  &   $0.242 \pm 0.034$ & $96.8 \pm 8.7$ & $0.935 \pm 0.095$ & $6.945 \pm 0.130$
\botrule
    \end{tabular}
    \begin{tabnote}
    {Input parameter vector $\theta=(0.2, 0.016, 0.15, 7.43, 1.9, 2.96, 1.4, 11.2)$
 corresponding to source with
$X=0.20$, $\zcno=0.016$, $Q_b=0.15$, $d=7.43$~kpc, $\xi_b=1.9$, $\xi_p=2.96$, $M_{\rm NS}=1.4\ M_\odot$, $R_{\rm NS}=11.2$~km, giving $1+z=1.259$.}\tnp
    \end{tabnote}
\end{table*}

We performed several experiments to attempt to replicate the simulated data via {\sc beansp}, varying the data treatment (see Table \ref{tab:runs}); including or omitting  the $\alpha$ values from the calculation; and fixing or allowing the neutron star $\mns$, $\rns$ to be free; and adding a prior on the $\xi_b$, $\xi_p$ values. 
In Table \ref{tab:sim_results} we list the inferred system properties against the input values.

\begin{figure*}
  \includegraphics{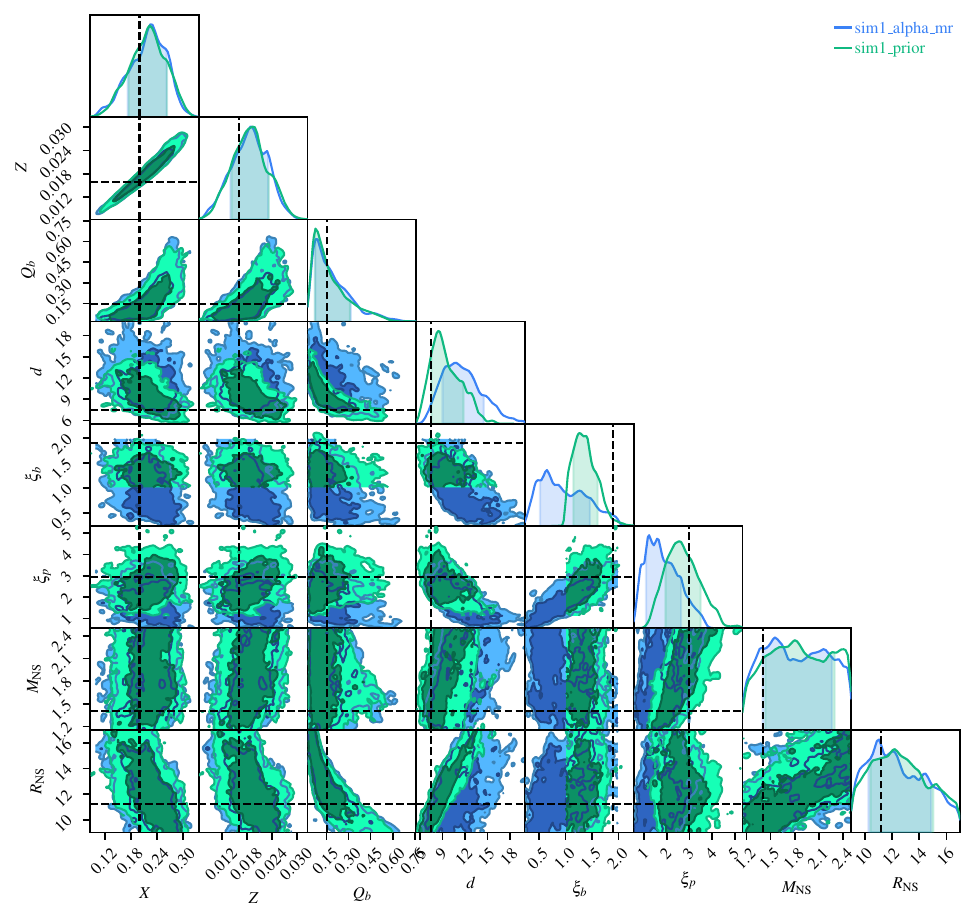}
 \caption{Comparison of posterior distributions for runs {\tt sim1\_alpha\_mr} and {\tt sim1\_prior}. The values of the model parameter input vector $\theta=(0.2, 0.016, 0.15, 7.43, 1.9, 2.96, 1.4, 11.2)$ are indicated by the black dashed lines in each panel. Note how the choice of a prior on the $\xi_b$, $\xi_p$ values also affects the distance $d$, but not the other model parameters. Other details as for Figure \ref{fig:base_sel_posteriors}}
  \label{fig:sim1_alpha_mr}
\end{figure*}

Each of our simulation runs consistently replicated the burst ignition properties, including the H-fraction $X$, CNO mass fraction \zcno, and the base flux $Q_b$, irrespective of the data treatment (Table \ref{tab:sim_results}).  However, each of the runs overestimated the distance $d$, and underestimated the anisotropy fractions $\xi_b$, $\xi_p$. Examples are shown in 
Figure \ref{fig:sim1_alpha_mr}, for the runs {\tt sim1\_alpha\_mr} and {\tt sim1\_prior}.

\begin{figure}
  \includegraphics{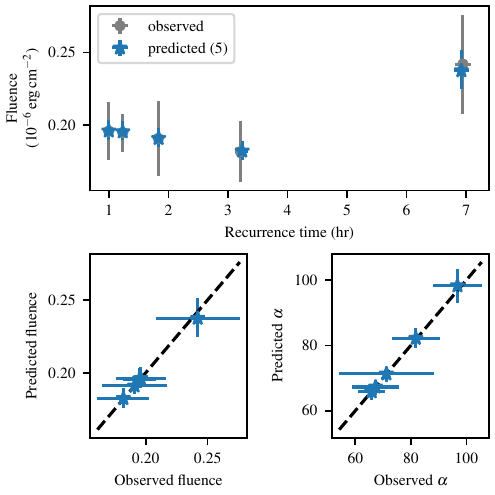}
 \caption{Comparison of predicted and input (simulated) burst properties for run {\tt sim1\_alpha\_mr}. The predictions in the top panel ({\it blue stars}) overlay (and hence obscure) the target measurement values. Other details as for Figure \ref{fig:base_sel_comparison}.}
  \label{fig:sim1_comparison}
\end{figure}

We evaluated the model on the last $10^5$ samples in the run and calculated the mean and $1\sigma$ variation in the predicted burst parameters, and 
{ show them compared to} the input values in Figure \ref{fig:sim1_comparison}. We found on average excellent agreement between the model predicted values for the recurrence time, fluence, and $\alpha$. The \ac{RMS} errors between the predicted and simulated recurrence times were an order of magnitude smaller than for the observations.
The variance in the model-predicted values was also substantially less than the uncertainty in the measured parameters.

The distance was overestimated for each of the runs, most dramatically for the run with $\mns$, $\rns$ free ({\tt sim\_alpha\_mr}). This deviation from the model input values was reflected in the anisotropy parameters $\xi_b$, $\xi_p$, which were underestimated in each case. The posterior 
{ distributions provide} 
only weak constraints on these parameters overall, such that underestimates in the anisotropy parameters lead to overestimates in the distance $d$.
To test whether more informative priors would help constrain these quantities better, we performed an additional run {\tt sim1\_prior} 
with a custom prior intended to restrict the anisotropy parameters $\xi_b$, $\xi_p$ to a range around the true (model input) value; but this effort was not completely successful, albeit bringing the distance and uncertainty down again to values that were consistent with the input value.

Two of the four runs performed were with $M$, $R$ fixed at $1.4\ M_\odot$, $11.2$~km respectively; for run {\tt sim1\_alpha\_mr} and {\tt sim1\_prior} we allowed these additional parameters to vary, but found no strong constraints as a result of the simulation data. We note that in both cases the $\mns$ posterior is centred on substantially higher values than the input, although with relatively large uncertainty. Thus, we conclude that the ensemble model comparison is broadly successful in accurately constraining the model parameters relating to burst ignition, with or without including the $\alpha$ values in the likelihood calculation; but less so for the system parameters including distance, anisotropy and neutron star mass and radius, particularly in the absence of an informative prior on these parameters.

\begin{table*}[b!]
 \caption{Inferred properties for the simulated data for \srga\ as listed in Table \ref{tab:sim_epochs}, for comparison with the model input values (column 2). Priors are as for the runs on the observational data, excluding the run {\tt sim1\_prior} which uses the functional prior on $\xi_b$, $\xi_p$ as described in the text. Other details as for Table \ref{tab:srga_results}.}\label{tab:sim_results}
\begin{tabular}{cccccccc}
\toprule
Parameter & Input value & Units  & {\tt sim1 } & {\tt sim1\_alpha} & {\tt sim1\_alpha\_mr} & {\tt sim1\_prior}  \\
\hline
$X$ & 0.20  &  & $0.21\pm0.03$ & $0.21\pm0.03$ & $0.22_{-0.05}^{+0.04}$ & $0.23_{-0.05}^{+0.04}$ \\
\zcno & 0.016 & & $0.016\pm0.003$ & $0.017\pm0.003$ & $0.019_{-0.005}^{+0.004}$ & $0.019\pm0.005$ \\
$Q_{\rm b}$ & 0.15  & MeV/nucleon & $0.16_{-0.04}^{+0.05}$ & $0.17_{-0.04}^{+0.06}$ & $0.16_{-0.09}^{+0.15}$ & $0.17_{-0.10}^{+0.14}$ \\
$d$ & { 7.43} & kpc & $8.9_{-1.1}^{+2.7}$ & $8.8_{-1.1}^{+2.9}$ & $11_{-2}^{+3}$ & $8.8_{-1.8}^{+2.6}$ \\
$\xi_b$ & 1.9 & & $1.3_{-0.6}^{+0.4}$ & $1.3_{-0.6}^{+0.4}$ & $0.9_{-0.4}^{+0.6}$ & $1.41_{-0.20}^{+0.29}$ \\
$\xi_p$ & 2.96 & & $2.1_{-0.8}^{+0.6}$ & $2.0_{-0.8}^{+0.6}$ & $1.7_{-0.6}^{+0.9}$ & $2.8_{-0.7}^{+1.1}$ \\
$M_{\rm NS}$ & 1.4 & $M_\odot$ & (fixed) & (fixed) & $1.8\pm0.4$ & $1.9\pm0.4$ \\
$R_{\rm NS}$ & 11.2 & km & (fixed) & (fixed) & $12.2_{-1.9}^{+2.7}$ & $12.1_{-1.8}^{+2.7}$ \\
\hline
RMS & & hr & { 0.00583} & { 0.0149} & { 0.0155} & { 0.0150}
\botrule
    \end{tabular}
\end{table*}

\subsection{\sax\ burst train mode}
\label{subsec:kepler}

We performed two {\sc beansp} runs on the simulated \sax\ {\sc kepler} data as described in Section \ref{subsec:crosscal}, with similar conditions as for the actual data as described in appendix \ref{sec:alphas} and \cite{gal24}.
The two runs differed only in the treatment of \mns and \rns, free or fixed (Table \ref{tab:runs}).

The {\sc kepler} input parameter were recovered reasonably well (Table \ref{tab:s1808_params}).
The input $Q_b$ value was within the $1\sigma$ confidence interval, but we suggest this is coincidence, as due to the different architecture of the two models ({\sc kepler} and {\sc settle}), we do not really expect that these parameters should correspond.

\begin{figure*}[ht!]
    \begin{centering}
        \includegraphics[width=\linewidth]{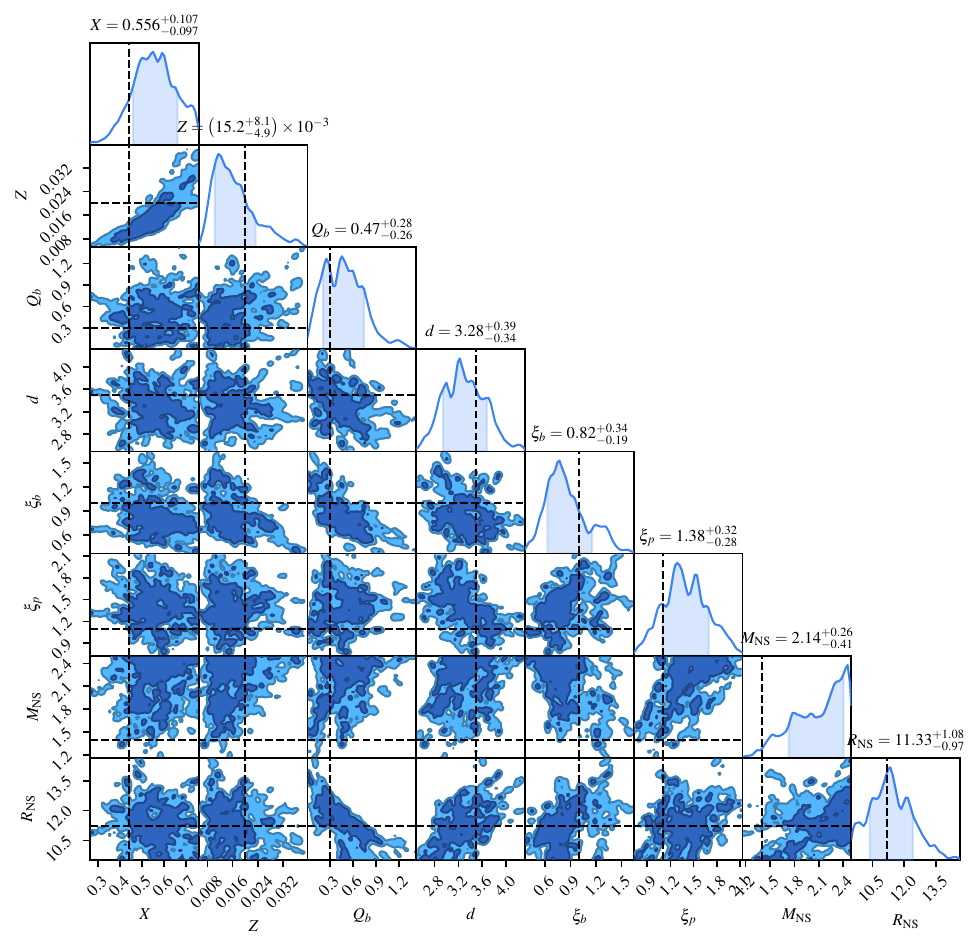}
        \caption{
            Neutron-star parameter posterior distributions for $10^5$ samples of the {\tt johnston18} simulation run to match {\sc kepler} predictions for \sax, with mass and radius free. The black dashed lines in each panel show the input parameters. Contours are at 1 and $2\sigma$.
        }
        \label{fig:johnston18}
    \end{centering}
\end{figure*}

Perhaps most interestingly, the neutron star radius was recovered well but the neutron star mass was significantly higher than the input value, approaching the prior limit (Figure \ref{fig:johnston18}). Consequently the surface gravity $g$ and redshift $1+z$ were all overestimated.
The posterior distributions demonstrated similar qualitative behaviour as for the observed data \cite[]{goodwin19c}, including the correlation between $X$ and $\zcno$.

\begin{table*}
	\caption{System parameters derived from the comparison with {\sc kepler} simulations for \sax. Other details as for Table \ref{tab:sim_results}.}
	\label{tab:s1808_params}
	\begin{tabular}{ccccc} 
		\hline
  Parameter & Input value & Units & {\tt johnston18c} & {\tt johnston18} \\
		\hline 
$X$ & 0.44        & & $0.51\pm0.14$ & $0.56_{-0.10}^{+0.11}$\\
\zcno & 0.02        & & $0.012_{-0.005}^{+0.009}$ & $0.015_{-0.005}^{+0.008}$\\
$Q_{\rm b}$ & 0.3 & MeV nucleon$^{-1}$ & $0.16\pm0.04$ & $0.5\pm0.3$\\
$d$ & 3.5         & kpc & $3.2_{-0.4}^{+0.5}$ & $3.3_{-0.3}^{+0.4}$\\
$\xi_b$ & (1.0)   & & $1.2\pm0.3$ & $0.82_{-0.19}^{+0.34}$ \\
$\xi_p$ & 1.1     & & $1.1_{-0.2}^{+0.3}$ & $1.4\pm0.3$\\
\mns &  1.4        & $M_\odot$ & (fixed) & $2.2_{-0.4}^{+0.3}$\\
\rns & 11.2        & km & (fixed) & $11.3_{-1.0}^{+1.1}$\\
\hline
RMS & & hr & { 0.0239} & { 0.0226} \\
		\hline
    \end{tabular}
\end{table*}

We caution that these results should be viewed in the strictest sense as a model comparison only, since \cite{johnston18} did not perform a comprehensive search of parameter space. That is, the {\sc kepler} simulations are based on a set of parameters which offer reasonable agreement to the observations, but do not necessarily represent the actual parameters for \sax. However, they 
lend further confidence of the capability of {\sc beansp} with {\sc settle} to recover system parameters with reasonable accuracy despite the simplicity of the ignition model.

\end{document}